\documentclass[useAMS,usenatbib]{mn2e}
\usepackage{graphicx}
\usepackage{float}
\usepackage{color}
\def\aap{Astronomy \& Astrophysics}
\def\aj{The Astronomical Journal}
\def\apj{The Astrophysical Journal}
\def\apjl{The Astrophysical Journal Letters}

\def\mnras{Monthly Notices of the Royal Astronomical Society}
\def\pasa{Publications of the Astronomical Society of Australia}
\def\pasp{Publications of the Astronomical Society of the Pacific}
\def\pasj{Publications of the Astronomical Society of Japan}
\def\nat{Nature}
\usepackage[switch]{lineno}
\title[A stellar over-density close to the SMC]{A stellar over-density associated with the Small Magellanic Cloud}

\author[Pieres,~A. et al.]{A. Pieres\thanks{E-mail: adriano.pieres@ufrgs.br}$^{1,2}$, B. X. Santiago$^{1,2}$, A.~Drlica-Wagner$^{3}$, K.~Bechtol$^{4}$, 
\newauthor
R.~P.~van~der~Marel$^{5}$, G.~Besla$^{6}$, 
N.~F.~Martin$^{7,8}$, V.~Belokurov$^{9}$, C.~Gallart$^{10,11}$, 
\newauthor
D.~Martinez-Delgado$^{12}$, 
J.~Marshall$^{13}$, N.~E.~D.~N{\"o}el$^{14}$, S.~R.~Majewski$^{15}$, 
\newauthor
M.-R.~L.~Cioni$^{16,17}$, T.~S.~Li$^{3}$, W.~Hartley$^{18}$, E.~Luque$^{1,2}$, B.~C.~Conn$^{19}$, 
\newauthor
A.~R.~Walker$^{20}$, E.~Balbinot$^{14}$, G.~S.~Stringfellow$^{21}$, K.~A.~G.~Olsen$^{22}$, 
\newauthor
D.~Nidever$^{22}$, L.~N.~da Costa$^{2,23}$, R.~Ogando$^{2,23}$, M.~Maia$^{2,23}$, A.~Fausti Neto$^{2}$,  
\newauthor
T.~M.~C.~Abbott$^{20}$, F.~B.~Abdalla$^{24,25}$, S.~Allam$^3$, J.~Annis$^{3}$, A.~Benoit-L{\'e}vy$^{25,26,27}$, 
\newauthor
A.~Carnero~Rosell$^{2,23}$, M.~Carrasco~Kind$^{28,29}$, J.~Carretero$^{30,31}$, C.~E.~Cunha$^{32}$, 
\newauthor
C.~B.~D'Andrea$^{33,34}$, S.~Desai$^{35}$, H.~T.~Diehl$^{3}$, P.~Doel$^{25}$, B.~Flaugher$^{3}$, 
\newauthor
P.~Fosalba$^{30}$, J.~Garc\'ia-Bellido$^{31}$, D.~Gruen$^{32,36}$, R.~A.~Gruendl$^{28,29}$, 
\newauthor
J.~Gschwend$^{2,23}$, G.~Gutierrez$^{3}$, K.~Honscheid$^{37,38}$, D.~James$^{20,39}$, K.~Kuehn$^{40}$, 
\newauthor
N.~Kuropatkin$^{3}$, F.~Menanteau$^{28,29}$, R.~Miquel$^{41,42}$, A.~A.~Plazas$^{43}$, A.~K.~Romer$^{44}$, 
\newauthor
M.~Sako$^{45}$, E.~Sanchez$^{46}$, V.~Scarpine$^{3}$, M.~Schubnell$^{47}$, I.~Sevilla-Noarbe$^{46}$, 
\newauthor
R.~C.~Smith$^{20}$, M.~Soares-Santos$^{3}$, F.~Sobreira$^{2,48}$, E.~Suchyta$^{49}$,
\newauthor
M.~E.~C.~Swanson$^{29}$, G.~Tarle$^{47}$, D.~L.~Tucker$^{3}$, W.~Wester$^{3}$ \\
Affiliations are listed after the references}
\begin{document}

\date{Accepted 2017 February, 23. Received 2017 February, 23; in original form 2016 December, 12}

\pagerange{\pageref{firstpage}--\pageref{lastpage}} \pubyear{2017}

\maketitle

\label{firstpage}

\begin{abstract}

We report the discovery of a stellar over-density 8$^{\circ}$ north of the center of the Small Magellanic Cloud (Small Magellanic Cloud Northern Over-Density; SMCNOD) using data from the first two years of the Dark Energy Survey (DES) and the first year of the MAGellanic SatelLITEs Survey (MagLiteS). The SMCNOD is indistinguishable in age, metallicity and distance from the nearby SMC stars, being primarly composed of intermediate-age stars (6 Gyr, Z=0.001), with a small fraction of young stars (1 Gyr, Z=0.01). The SMCNOD has an elongated shape with an ellipticity of 0.6 and a size of $\sim$ 6x2 deg. It has an absolute magnitude of $M_V \cong$ -7.7, $r_h = 2.1$ kpc, and $\mu_V(r<r_h)$ = 31.2 mag arcsec$^{-2}$. We estimate a stellar mass of $\sim 10^5$ $M_{\odot}$, following a Kroupa mass function. The SMCNOD was probably removed from the SMC disk by tidal stripping, since it is located near the head of the Magellanic Stream, and the literature indicates likely recent LMC-SMC encounters. This scenario is supported by the lack of significant HI gas. Other potential scenarios for the SMCNOD origin are a transient over-density within the SMC tidal radius or a primordial SMC satellite in advanced stage of disruption.
\end{abstract}


\section{Introduction}
\label{intro}
The Magellanic Clouds (MCs) are a rich and nearby system where we can observe dynamic evolution as well as the results of star formation throughout time. The system also includes the Magellanic Stream (MS), a HI gas stream~\citep{1974ApJ...190..291M} connected to the MCs spanning at least 200$^{\circ}$ on the sky~\citep{2010ApJ...723.1618N}, where no stellar counterpart has yet been identified~\citep{1982MNRAS.201..473R,1998ASPC..136...22G}. Other important structures belonging to this system are the Magellanic Bridge, containing neutral hydrogen, stars and star clusters linking the MCs~\citep{2015MNRAS.453.3190B,1992AJ....103.1234G,1990AJ.....99..191I} and the Leading Arm (LA) or Leading Arm Feature (LAF), a gas stream on the opposite side of the MS.

Given the higher velocities (than previously estimated) for the MCs in recent works~\citep{2013ApJ...764..161K,2010AJ....140.1934V,2006ApJ...652.1213K,2006ApJ...638..772K}, it is {thought} that the MCs are completing their first passage around the Milky-Way (MW). This conclusion is supported by proper motion measurements using HST \citep{2013ApJ...764..161K} and \textit{Gaia} data release 1~\citep{2016ApJ...832L..23V}. Thus, the gravitational interaction between the Small and the Large Magellanic Clouds (SMC and LMC, respectively) may be playing a larger role than the MW in triggering star formation. 

In the recent decades, a wide range of dynamical simulations of the MCs have improved our understanding of their substructures, taking advantage of more reliable proper motion measurements, among other enhanced initial conditions (e. g., masses, gas fraction, ellipticity, stellar disk scale length). Using N-body simulations, 
\citet{2006MNRAS.371..108C} reproduced the MS and LA as substructures formed through tidal interaction between the Clouds. Their work reproduced for the first time the spatial and kinematic bifurcations in the LA and in the MS. The MCs simulations of \citet{2007PASA...24...21B} over the last 800 Myr are able to reproduce the off-center bar and the HI spirals in the LMC. They also predict that a substantial number of SMC stars could be transferred to the LMC to form diffuse halo components around that galaxy.
Restricting their study to the SMC, \citet{2009PASA...26...48B} designed chemodynamical simulations using a SMC `dwarf spheroidal model' (an extended HI gas disk and a spherical distribution for old stars), which they argue is a better description of the stellar and gas kinematic properties. In their fiducial model, the final distribution of old stars is more regular (spherical) than that of the younger stars (which form basically a bar-like structure).
~\citet{2012ApJ...750...36D} simulate a large set of models based on proper motion data from~\citet{2010AJ....140.1934V} and from~\citet{2006ApJ...652.1213K,2006ApJ...638..772K}, predicting two main encounters between the SMC and the LMC (260 Myr and 1.97 Gyr ago), suggesting a joint history for these galaxies. In their simulations, the first encounter forms two substructures: the Magellanic Bridge and a less obvious structure called Counter-Bridge.
\citet{2012MNRAS.421.2109B} present two models for the Magellanic System, designed to explain the MS as the action of LMC tides on the SMC. In their models the LMC is a one-armed spiral and features as well a warped, off-centre stellar bar as a result of the gravitational interaction.

The possible association~between the MCs and ultra-faint dwarf galaxies recently discovered in the Dark Energy Survey~\citep[DES;][]{2005astro.ph.10346T} footprint has revived the search for dwarf galaxy satellites of the LMC or SMC~\citep{2016arXiv160902148D,2016arXiv160503574S,2016arXiv160304420J,2015ApJ...813..109D,2015ApJ...805..130K,2015MNRAS.453.3568D}.  The recent discovery by~\citet{2016MNRAS.459..239M} of a stellar cloud with a length of 10 kpc within the LMC tidal radius (and an additional extension farther west of the LMC) shows that the exploration of the outer area of the MCs has an important potential for new discoveries. It also reinforces the idea that newly discovered structures can be used to trace the gravitational interaction history of the MCs.~\citet{2016ApJ...825...20B} suggest that the existence of stellar arcs and multiple spiral arms in the northern LMC periphery (without comparable counterparts in southern regions of the SMC) could be attributed to repeated close interactions between the LMC and the SMC. A large number of simulations predict clumpy substructures formed by a spheroidal distribution surrounding the SMC (see for example the references listed in Section~\ref{origin}), though there is no specific prediction of over-densities as large as those presented here. Nevertheless, the discovery of this structure reinforces the scenario where the LMC and the SMC have had recent and drastic encounters.

In what follows, we report a stellar over-density located 8$^{\circ}$ north of the SMC center, hereafter referred to as the Small Magellanic Cloud Northern Over-Density (SMCNOD). The SMCNOD was discovered in data from DES and follow up imaging was performed with the Dark Energy Camera~\citep[DECam;][]{2015AJ....150..150F} as part of the MAGellanic SatelLITEs Survey - MagLiteS. The data sets and criteria used to select stellar sources are discussed
in Section~\ref{data}. In Section~\ref{analysis} we describe the analysis of the stellar populations and the structure of the SMCNOD. We conclude by discussing the SMCNOD stellar population and gas content, as well as its formation and fate, in Section~\ref{discussion}.

\section{Data}
\label{data}

% We present a brief description of the data set used here from both DES and MagLiteS observations.

\begin{figure*}
 \includegraphics[width=480px]{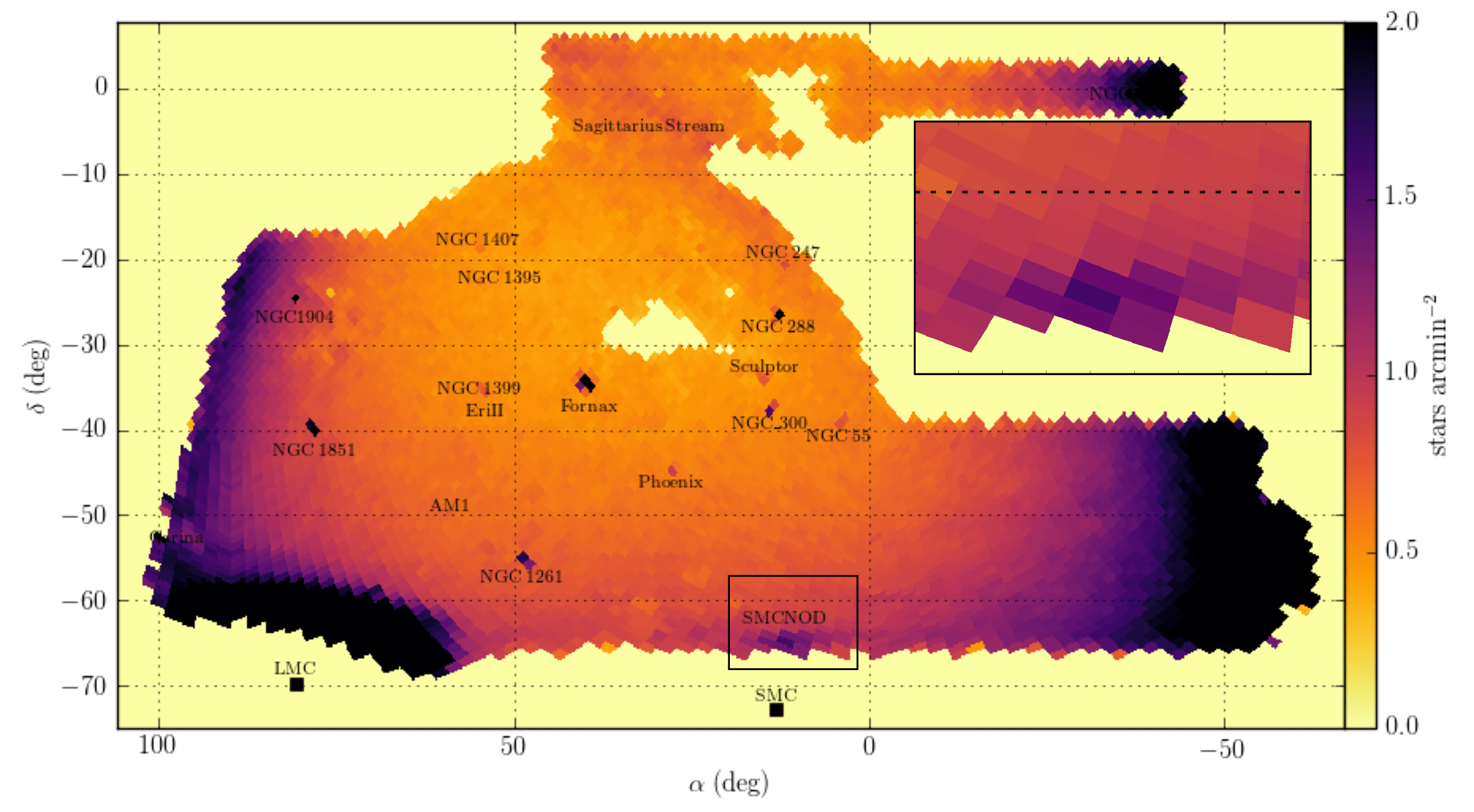}
 \caption{DES-Y2Q1 stellar density map using the~\textsc{HealPix} scheme with~\textsc{nside} = 64 (0.839 square degrees), corrected by a survey coverage map and saturating at 2 stars arcmin$^{-2}$. Objects detected in the search for extended structures are explicitly labeled in the DES-Y2Q1 footprint. In this density map we are counting all of the DES-Y2Q1 stars.~\textit{Inserted box:} Zoomed view of the SMCNOD region.}
\label{DM23}
\end{figure*}

The DES data used in this work is the year-two quick release (DES-Y2Q1) catalog, constructed using 26,590 DECam exposures. The DES-Y2Q1 images were taken between August 2013 and February 2014 and between August 2014 and February 2015, in the first two years of the survey. The images cover most of the DES footprint (5,000 square degrees), with the exception of a few hundred square degrees in the region near the South Galactic Pole. We refer to~Section 2 of~\citet{2015ApJ...813..109D} for a detailed description of the data and the star selection criteria.

MagLiteS is a National Optical Astronomy Observatories (NOAO) community survey (NOAO proposal 2016A-0366) that is using the DECam to complete an annulus of contiguous imaging around the periphery of the Magellanic System~\citep{2016arXiv160902148D}.

The MagLiteS data used here are composed of 16 DECam 90s exposures in the $g$ and $r$ bands. The positions for each MagLiteS DECam exposure are listed in Tab.~\ref{pos}. The MagLiteS exposures were taken in 27 June 2016, in an effort to enlarge DECam coverage in the SMCNOD region. MagLiteS images were reduced using the Dark Energy Survey Data Management (DESDM) pipeline, and source detection was performed separately on each exposure.

\begin{table}
\begin{center}
\begin{tabular}{ l | c | c }
\hline
Band & $\alpha (^{\circ})$ & $\delta (^{\circ})$\\
\hline
$g$ & 14.092 & -65.826 \\
  & 10.690 & -65.029 \\
  & 7.280 & -65.861 \\
  & 10.804 & -66.803 \\
  & 9.789 & -65.893 \\
  & 13.324 & -66.744 \\
  & 16.504 & -65.664 \\
  & 13.055 & -64.994 \\
$r$ & 14.114 & -65.815 \\
  & 10.699 & -65.036 \\
  & 7.306 & -65.856 \\
  & 10.807 & -66.783 \\
  & 9.792 & -65.893 \\
  & 13.322 & -66.729 \\
  & 16.524 & -65.653 \\
  & 13.077 & -65.004 \\
\hline
\end{tabular}
\caption{List of bands and central positions for each MagLiteS DECam exposures used in this work.}
\label{pos}
\end{center}
\end{table}

To  assemble a combined DES-Y2Q1 and MagLiteS source catalog, we first set the zero-points by comparing DES-Y2Q1 bright stars to individual MagLiteS DECam single-exposure catalogs. Since the DES-Y2Q1 is a de-reddened catalog, we applied an extinction correction to each DES-Y2Q1 source following~\citet{1998ApJ...500..525S}. Comparison stars were selected in the magnitude range of $17 < \emph{g} < 21$, $\mid\!wavg\_spread\_model\!\mid\ < 0.003$ and $flags < 4$ in each band ($g$ and $r$) in the DES-Y2Q1 catalog. \textsc{spread\_model}\footnote{ \textsc{spread\_model} is a ``normalized simplified linear discriminant between the best-fitting local PSF model and a slightly more extended model'' as described in~\citet{2012ApJ...757...83D}.} is a morphological output from \textsc{SExtractor}\footnote{http://www.astromatic.net/software/sextractor} used to distinguish stars from galaxies. The prefix \textit{wavg} means we used the weighted average of \textit{spread\_model} from individual single-epoch detections. The maximum positional deviation (object matching between DES-Y2Q1 and MagLiteS sources) was set to 1 arcsec. After adding the photometric zero points, we joined all sources from the MagLiteS fields into a single catalog. We then subtracted the extinction and incorporated the final MagLiteS catalog into the DES-Y2Q1 catalog, to create the final DES-MagLiteS stars list used in this paper. We applied the same criteria used to select DES-Y2Q1 stars to filter our final sample of stars, namely using a star/galaxy separation criterion of $\mid\!wavg\_spread\_model\_r\!\mid < 0.003+spreaderr\_model\_r, flags\_\{g, r\} < 4$, and $magerr\_psf\_\{g, r\} < 1$.
%  made from the same PSF convolved with a circular exponential disk model with scale length = full width half maxima/16

Moreover, we applied a magnitude cut of $17 < \emph{g} < 23$ to ensure high source detection efficiency on the DES-MagLiteS catalog. Also, we applied a color cut to select stellar sources with $-0.5 <\emph{g-r} < 1.2$.

A single 90 s DECam exposure (the DECam exposure time for DES in~\emph{g} and~\emph{r} bands) reaches point sources with magnitudes as faint as \{$g$,$r$\} $\cong$ \{23.6,23.2\} with a signal to noise ratio (SNR) equal to 10. Therefore, the faint magnitude cut adopted here results in uniform depth at least down to this SNR level. We emphasize that the quoted limiting magnitudes and SNR may change slightly due to seeing and weather conditions during the observing nights.

\begin{figure*}
\centering
\includegraphics[width=480px]{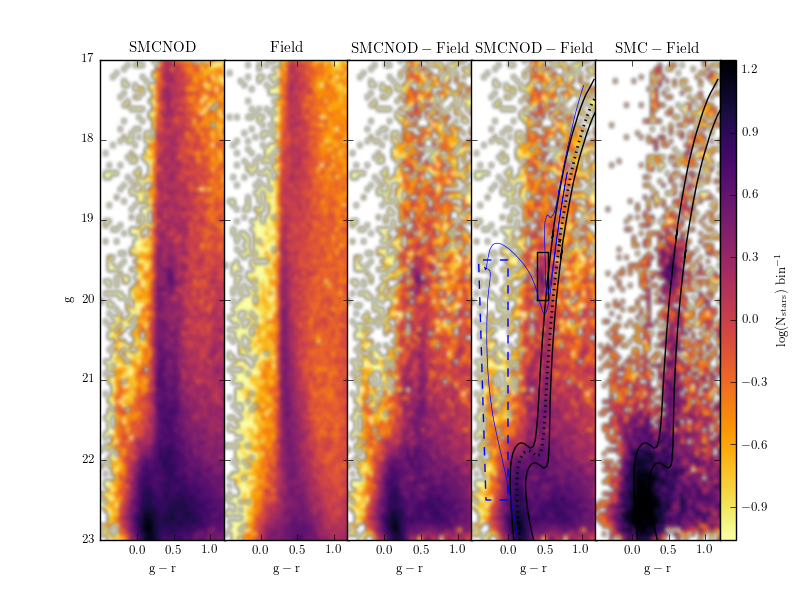}
\caption{\emph{First panel:} Hess diagram for stars in the range $6^{\circ}<\alpha<15^{\circ}$ and $-66^{\circ}<\delta<-63^{\circ}$, which covers most of the SMCNOD. A stellar population with a turnoff at~\emph{g} $\approx$ 22 can be clearly seen. A main sequence (MS), sub-giant branch (SGB), red giant branch (RGB) and red clump (RC) are all discernible against the foreground Galactic stars even with no subtraction. \emph{Second panel:} Hess diagram for a region at the same Galactic latitude (\emph{b} = -52.58$^{\circ}$) as the previous one ($32^{\circ}<\alpha<40^{\circ}$, $-64^{\circ}<\delta<-58.25^{\circ}$).~\emph{Third panel:} Subtracted Hess diagram (object minus field).~\emph{Fourth panel:} Same as the third panel, but featuring isochronal masks of intermediate-age (young) SMC populations bounded by the solid black (dashed blue) line, encompassing most of the SMCNOD stars by displacing the isochrone. The black rectangle denotes the RC stars. A PARSEC model with $\tau \simeq$ 6 Gyr and Z = 0.001 ($\tau \simeq$ 1 Gyr, Z = 0.01) is shown in dotted black (solid thin blue) line. Both models (intermediate-age and young) are displaced by a distance modulus equal to 18.96, following the SMC distance modulus obtained by~\citet{2015AJ....149..179D}. ~\emph{Fifth panel:} Sample of SMC field stars with $\delta < -67^{\circ}$ and $10^{\circ}< \alpha < 15 ^{\circ}$ minus the second Hess diagram (MW foreground stars). The data show a spread in magnitude that is comparable between the SMC stars and SMCNOD stars. The isochronal mask for intermediate-age stars is reproduced in the last panel.}
\label{Hessdiagram}
\end{figure*}

\section{Analysis}
\label{analysis}

The SMCNOD was discovered during a search for extended ($r_t \simeq$ 30 arcmin) and low surface brightness structures in the DES-Y2Q1 catalog. We initially built density maps for the DES-Y2Q1 stars, partitioning the sky into equal area \textsc{HealPix}\footnote{\textsc{HealPix} is an equal-area pixelization scheme for spherical surfaces (in our case, the sky) in an certain number of pixels. This number of pixels is given by 12 times the square of the parameter \textsc{nside}, chosen by the user. See more details in http://healpix.sourceforge.net/ }. We set the pixel area to 0.839 square degrees (\textsc{nside} = 64). We then counted all DES-Y2Q1 stars within each pixel, correcting the density in each pixel by the respective survey coverage for both~\emph{g} and~\emph{r} bands. The coverage maps were created by using a finer grid of pixels (pixel size $\cong$ 1.7 arcmin on a side, \textsc{nside} = 2048), then checking whether or not a pixel contains any star or galaxy and finally grouping into pixels with 0.839 square degrees (\textsc{nside} = 64), where the effective survey coverage area was computed. We then calculated the average number of stars in the 8 immediately neighboring pixels (with~\textsc{nside} = 64). The significance of any over-density was calculated by subtracting the average counts in the neighboring pixels and dividing the result by the square root of that average, thus yielding the number of standard deviations (following a Poisson distribution) of the star counts. For example, the least significant candidate has a star count equal to 3781, whereas the average counts of the neighboring pixels is 1.0 star per arcmin$^2$ (3600 stars per square degree). Its significance is then only 3$\sigma$ ($(3781-3600)/\sqrt[]{3600}$), presenting an excess of 5\% above the mean star counts. We examined all candidates with significance greater than 3$\sigma$, which results in a list of 314 candidates.

The highest significance candidates were mostly known globular clusters and dwarf galaxies (see Figure~\ref{DM23}). However, one candidate located at $\alpha \cong 12^{\circ}$ and $\delta \cong -65^{\circ}$ was significantly higher (with a significance of 8$\sigma$ at the highest density pixel) than the local background, and several of its neighboring pixels emerged in the significance list, suggesting that the over-density spans multiple pixels (insert in Figure~\ref{DM23}). Given its proximity to the SMC, we refer to this over-density as the SMCNOD. This object is located in the border of the DES survey and se we performed follow up imaging with MagLiteS to cover an extra area around SMCNOD.
\begin{figure*}
\centering
\includegraphics[width=500px]{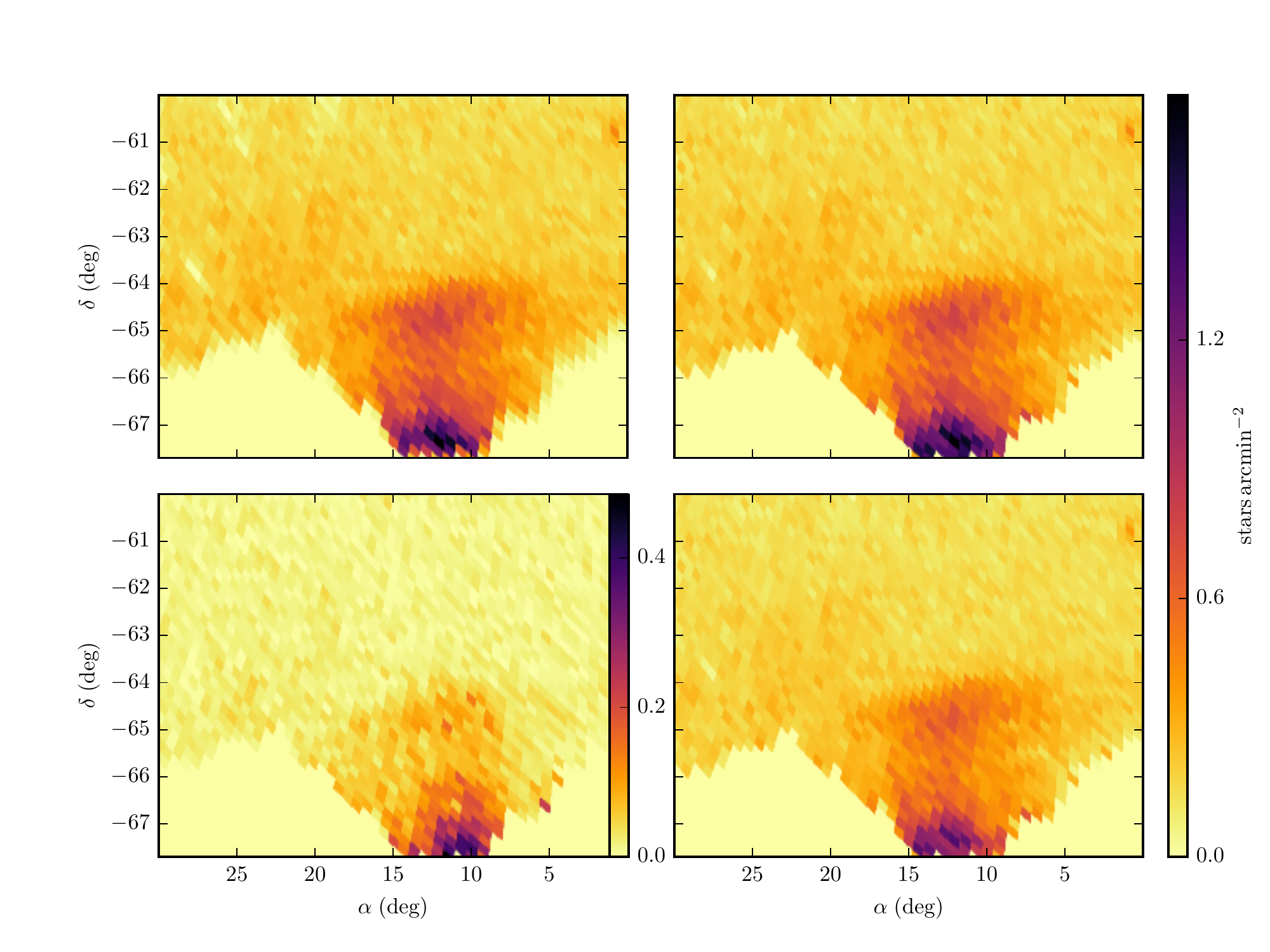}
\caption{~\emph{Top left:} Density map (each pixel has 14 $\times$ 14 arcmin, \textsc{nside}=256) for stars filtered by the isochrone mask shown in the fourth panel of Figure~\ref{Hessdiagram}, in the field surrounding the SMCNOD.~\emph{Top right:} The same density map as in the left, but corrected by the coverage map.~\emph{Bottom:} Density map for young (left) and intermediate-age (right) stellar population. Both bottom panels are corrected by the coverage map. The object in the top right corner of each panel at $\alpha,\delta \simeq \{1^{\circ}, -61^{\circ}\}$ is the dwarf galaxy candidate Tucana IV~\citep{2015ApJ...813..109D}. Top left panel and both right panels are sharing the rightmost colorbar, while the young population is shown in a different colorbar scale, to highlight its weak contribution.}
\label{DMcorr}
\end{figure*}

In the leftmost panel of Figure~\ref{Hessdiagram} we plot the \emph{g} vs. \emph{g-r} color-magnitude Hess diagram for the region surrounding the SMCNOD in the DES-MagLiteS catalog, to analyze the photometric features of that putative stellar population. The second Hess diagram samples stars in a field with 20 square degrees, centered on $l=304.60^{\circ}$ and $b=-52.60^{\circ}$ at the same Galactic latitude as the SMCNOD ($l=284.72^{\circ}$, $b=-52.60^{\circ}$). Subtracting the first two Hess diagrams (and weighting by their respective areas) results in the third Hess diagram in Figure~\ref{Hessdiagram}. It is dominated by a stellar population with age $\tau \simeq$ 6 Gyr and metallicity Z = 0.001 as attested by the overlaid PARSEC model~\citep{2012MNRAS.427..127B} represented by the dotted black line in the fourth panel. This model was chosen by a visual comparison to the Hess diagram in the fourth panel. A CMD mask is drawn (solid black lines) displacing the PARSEC model for intermediate-age stars in $g-r$ color and $g$ magnitude. The SMCNOD distance modulus is indistinguishable from that of the SMC ($18.96\pm0.02$ following~\citealt{2015AJ....149..179D}). A PARSEC model is also shown in fourth panel of Figure~\ref{Hessdiagram} (thin blue line) to represent the blue plume of younger stars ($\tau \simeq$ 1 Gyr and Z=0.01). We note there is some overlap between both populations (younger and intermediate-age) in the lower Main Sequence, Red Giant Branch (RGB) and Red Clump (RC) color-magnitude diagram (CMD) regions. In the last Hess diagram, stars from the DES-MagLiteS catalog with $-68^{\circ}<\delta<-67^{\circ}$ and $10^{\circ}<\alpha<15^{\circ}$ (the closest region to the SMC in the DES-MagLiteS catalog) are sampled and the CMD mask for intermediate-age stars is reproduced, to compare the SMC and SMCNOD stellar populations.

\begin{table*}
\begin{center}
\begin{tabular}{ l | c | c | c | c | c | c | c }
\hline
Field name & $\alpha\ (^{\circ})$ & $\delta\ (^{\circ})$ & $l\ (^{\circ})$ & $b\ (^{\circ})$ & $\rho$ stars degree$^{-2}$ & $\rho$ giants degree$^{-2}$\\
\hline
SMCNOOD center & 12.000 & -64.800 & 303.529 & -52.317 & 457 & 210 [166]$^{\dagger}$ & \\
84S341 & 6.892 & -64.741 & 307.082 & -52.194 & 357 & 110 [87.2]$^{\dagger\dagger}$ \\
MW foreground & 19.928 & -64.600 & 297.985 & -52.256 & 247 & -- \\
 \hline
 \multicolumn{7}{l}{$^{\dagger}$ Density of giants estimated for the SMCNOD center} \\
\multicolumn{7}{l}{$^{\dagger\dagger}$ Density of giants from~\citet{2011ApJ...733L..10N}, used as reference. Both ($^{\dagger}$ and $^{\dagger\dagger}$) are plotted in Fig.~\ref{Nid}.} \\
\end{tabular}
\caption{Name (first column) and position in equatorial (columns 2 and 3) and Galactic (columns 4 and 5) coordinates for three fields: the SMCNOD center, the field overlapping~\citet{2011ApJ...733L..10N} and a MW foreground field at roughly the same Galactic latitude. The numbers in the second to last column are the stellar density after applying our CMD filters. The last column presents the density of giants after subtracting the MW foreground density (247 stars degree$^{-2}$). Numbers in brackets in the last columns are normalizing to~\citet{2011ApJ...733L..10N}, used as reference. More details are in the text.}
\label{table}
\end{center}
\end{table*}

The PARSEC model for intermediate-age is a good description of the SMCNOD population, and we selected stars that are more likely to belong to the object using the CMD filter described above. As a young population is also visible in the third and fourth subtracted Hess Diagrams, we added an extra filter box to include the younger Main Sequence stars.  Using both CMD filters described above (for intermediate-age and young stars), we reanalyzed the stellar density distribution in the DES-MagLiteS catalog. First, we built the stellar density map (top left panel in Figure~\ref{DMcorr}) using \textsc{nside}=256 (pixel size $\cong$ 14 arcmin on a side). 
Dividing this stellar density map by the coverage map results in the stellar density map shown in top right panel in Figure~\ref{DMcorr}. We now see a stellar over-density with a roughly elliptical shape, mainly composed of intermediate-age stars (comparing both bottom panels of Figure~\ref{DMcorr}) at a distance of 8$^{\circ}$ from the SMC center ($\alpha = 13.000^{\circ}$, $\delta = -72.817^{\circ}$). 

We follow the model from~\citet{2007ApJ...665L..23N} to compare the SMCNOD brightness to the expected SMC surface brightness extrapolated to that position. They fit the SMC surface brightness profile (in $B$ and $R$ bands) using three 34 arcmin $\times$ 33 arcmin fields located southwards of the SMC, at a distance of 4.7, 5.6 and 6.5 kpc (respectively 4.2$^{\circ}$, 4.9$^{\circ}$ and 5.8$^{\circ}$). Extrapolating their surface brightness profile out to a radial distance of 8$^{\circ}$ from the SMC center, we derive an expected $B$ band surface brightness of $\mu_B$ = 32.4 $\pm$ 0.3 mag arcsec$^{-2}$. To compare to the SMCNOD surface brightness we first applied a transformation of stellar magnitudes from~\emph{g} and~\emph{r} DES bands to~\emph{g} and~\emph{r} SDSS bands, following~\citet{2015ApJ...807...50B}:

\begin{equation}
g_{SDSS} = g_{DES} + 0.104 (g_{DES}-r_{DES}) - 0.01
\end{equation}
\begin{equation}
r_{SDSS} = r_{DES} + 0.102 (g_{DES}-r_{DES}) - 0.02
\end{equation}
We then converted the SDSS magnitudes from the CMD filtered stars to the $B$ band using the transformation equation from~\citet{2005AJ....130..873J}:

\begin{equation}
B = g_{SDSS} + 0.390 (g_{SDSS}-r_{SDSS}) + 0.21
\end{equation}

We evaluate the integrated $B$ flux at the SMCNOD center, in the same \textsc{HealPix} pixels (\textsc{nside} = 256) applied before, obtaining a surface brightness of $\mu_B = 29.7 \pm 0.17$ mag arcsec$^{-2}$. This is almost three magnitudes brighter than expected from extrapolating the main body of the SMC based on~\citet{2007ApJ...665L..23N}. The uncertainties were estimated using a bootstrap method, where the stars in the central pixel were randomly sorted (with replacement) to make up a new estimate of the brightness in the B band. A total of 1000 such bootstrap realizations were carried out.

%\begin{figure}
%\centering
%\includegraphics[width=230px]{Bband.pdf}
%\caption{Density flux for B band in magnitudes arcsec$^{-2}$.}
%\label{Bband}
%\end{figure}

\begin{figure}
\centering
\includegraphics[width=240px]{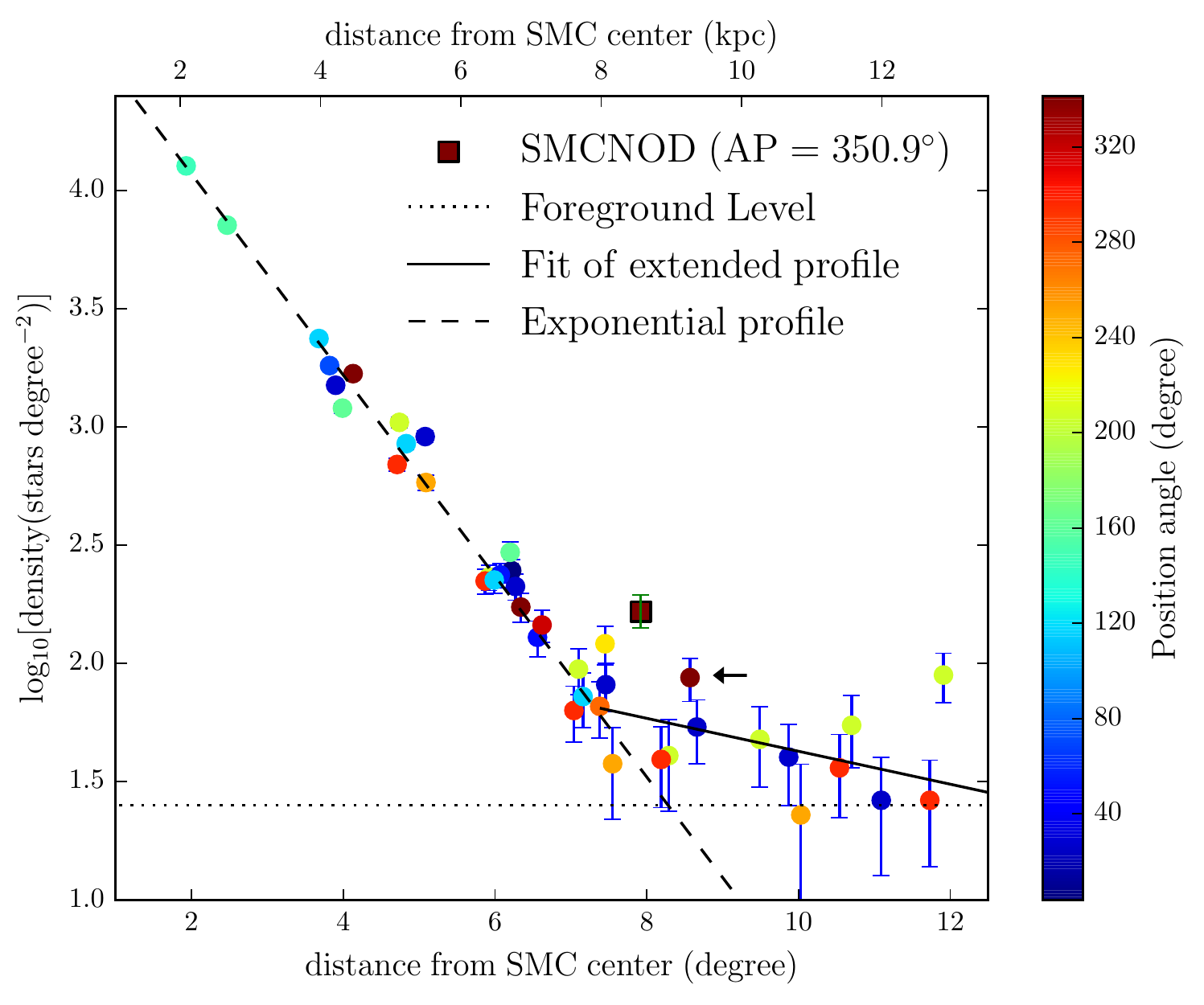}
\caption{RGB density profile reproduced from Fig. 3 of ~\citet{2011ApJ...733L..10N}. The circles are the densities (as found by those authors) color coded by position angle (from N to E). Also shown are the foreground contamination level and fitted profiles. The brown square is the density of giants as sampled in the SMCNOD center, while the brown circle at 8.4$^{\circ}$ (indicated by an black arrow) is the field 84S341, which overlaps with DES-MagLiteS and which was used to re-normalize DES-MagLiteS density scale. The SMC center adopted here is at $\alpha=15.129^{\circ}, \delta=-72.720^{\circ}$ and a distance of 61.94 kpc from the Sun~\citep{2015AJ....149..179D}.}
\label{Nid}
\end{figure}

\citet{2011ApJ...733L..10N} explored the SMC RGB distribution using data from the MAgellanic Periphery Survey (MAPS), sampling 36 arcmin $\times$ 36 arcmin fields with the MOSAIC II Camera mounted on the CTIO 4m Blanco telescope, and reaching as far as 12$^\circ$ from the SMC center. They observed stars with Washington photometry in three bands ($DDO51$, $M$ and $T_2$), as these bands are useful to discriminate MW foreground dwarfs from SMC RGB stars. The best-fit elliptical density profile for the SMC giants sampled presents a ``break'' at 7.5$^\circ$ from the fitted center ($\alpha=15.129^{\circ}, \delta=-72.720^{\circ}$), where the density slope abruptly decreases and the distribution of giants begins to scatter around this flatter profile. In Figure~\ref{Nid} we reproduce Figure 3 from~\citet{2011ApJ...733L..10N} using color coded circles according to the field position angles. The figure also reproduces the best-fit models both internal and external to the profile break. 

To compare the density of RGB stars at the center of the SMCNOD to the measurements of~\citet{2011ApJ...733L..10N} we subtracted the Galactic foreground dwarf stars contaminating the RGB locus and we normalized our densities to the density profile shown in Figure~\ref{Nid}. This second goal is achieved with the use of the only field from~\citet{2011ApJ...733L..10N} that overlaps the DES-MagLiteS footprint, which they name 84S341 and which is located 2.2$^\circ$ away from the SMCNOD center. Table~\ref{table} lists the positions and DES-MagLiteS stellar densities at the SMCNOD center, at the 84S341 field and at a field far from the SMCNOD. The three fields are at nearly the same Galactic latitude and we assume that the difference in MW dwarf counts are negligible. The densities listed in the second to last column correspond to the stars falling within the intermediate-age isochrone CMD mask described earlier with an additional color-magnitude cut of $g<$ 21 applied and with no RC stars included. We refer to that filter as the RGB box. The first~line in the last column represents the resulting RGB density after subtracting the foreground contamination. The bracketed density value for the 84S341 field is the RGB density actually measured by~\citet{2011ApJ...733L..10N}. The final SMCNOD RGB density (also shown in brackets) is then obtained by applying the same ratio as in the 84S341 field (166 giants degree$^{-2}$), placing it clearly above the density profile of any of the individual fields analyzed by~\citet{2011ApJ...733L..10N} at that angular distance (Figure~\ref{Nid}). 

\begin{figure}
\centering
\includegraphics[width=260px]{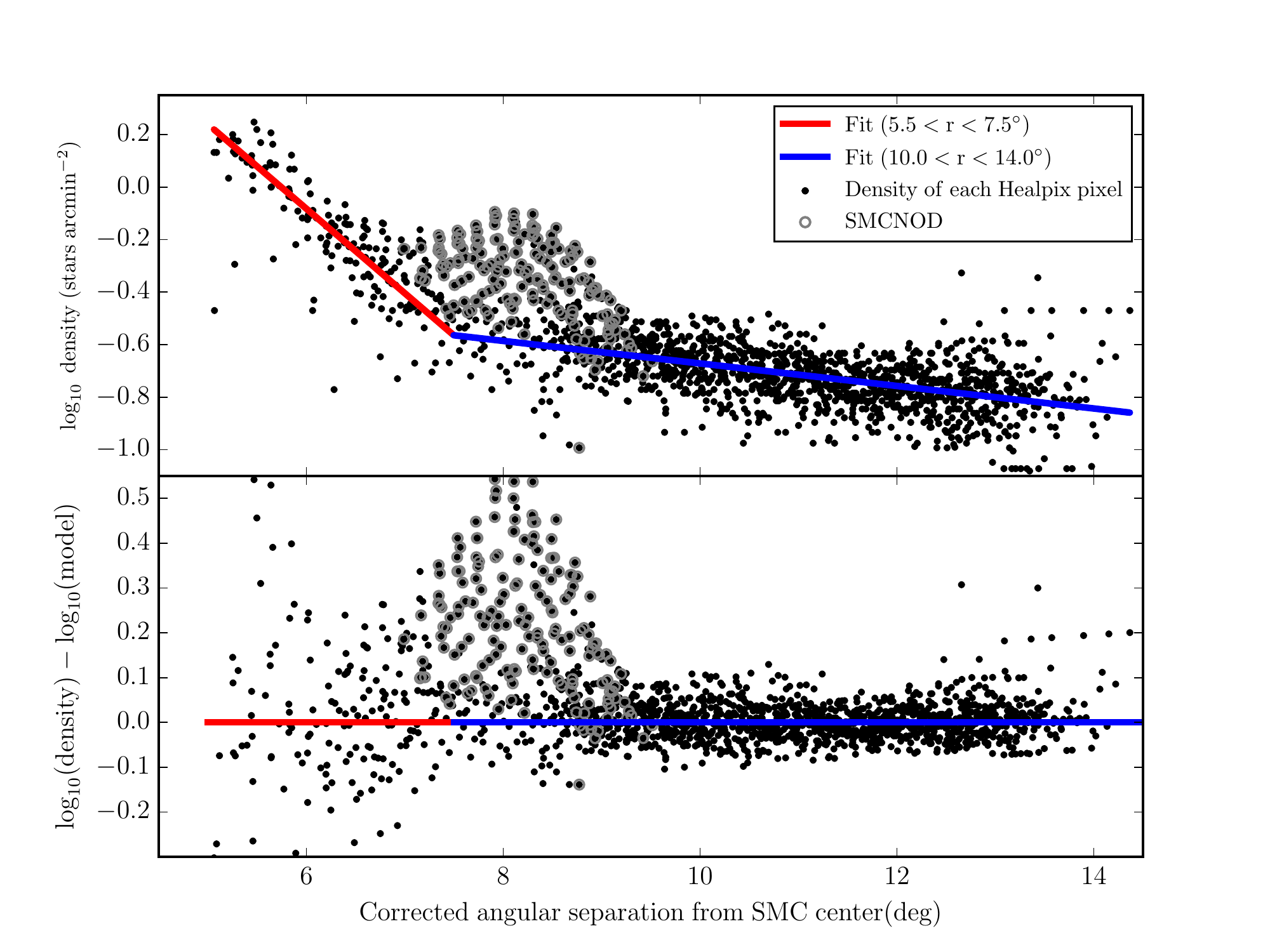}
\caption{~\emph{Top}: Stellar density (black dots) versus angular separation (corrected for elliptical shape and position angle) of the SMC center. The stellar density was determined in each~\textsc{HealPix} (\textsc{nside}=256) in DES-MagLiteS, and the angular separation corresponds to the elliptical exponential model from~\citet{2011ApJ...733L..10N}. The solid red line is the fit for~\textsc{HealPix} between 5.5$^{\circ}$ and 7.5$^{\circ}$ and solid blue line for pixels between 10.0$^{\circ}$ and 14.0$^{\circ}$. The grey circles are the boxes within the SMCNOD position (cells within truncation radius).~\emph{Bottom:} Stellar density data divided by the model for all~\textsc{HealPix} pixels. The SMCNOD resides at the interface between exponential models, but is discrepant from both. The radial scale length is 1.33$^{\circ}$ for the inner fit and 10.13$^{\circ}$ for the outer fit.}
\label{DPC}
\end{figure}

To compare the SMCNOD stellar density to the surrounding areas we fit two models: a profile closer to the SMC than the SMCNOD (called the inner profile) and a profile more distant of the SMC than the SMCNOD (the outer profile). The distances from the SMC center to each~\textsc{Healpix} pixel were set following the~\citet{2011ApJ...733L..10N} elliptical exponential model. The density of CMD-filtered stars were calculated in~\textsc{HealPix} pixels with \textsc{nside} = 256. We fitted the inner (outer) profile for boxes between 5.5$^{\circ}$ and 7.5$^{\circ}$ (between 10$^{\circ}$ and 14$^{\circ}$) from the SMC center. The fits provide an independent and striking confirmation of the SMC extended profile, along with the break at $\simeq 8^{\circ}$ from the SMC center. The top panel in Figure~\ref{DPC} shows both inner (red line) and outer (blue line) fits for the density profiles near the SMCNOD. Dividing the density by the respective fits (bottom panel in Figure~\ref{DPC}), we find that the~\textsc{HealPix} pixels within the SMCNOD truncation radius (where the densities decrease to the level of the background density) have notably higher densities than those of the surrounding areas. %The density profile for old population (6 Gyr) follows roughly the same profile accounting both (young and old) population, while the young stars present a~sparsely sampled} profile (typical densities (0.1 stars arcmin$^{-2}$ out to 9$^{\circ}$). As seen in Figure~\ref{DMcorr}, the SMCNOD is dominated by the intermediate-age population (6 Gyr), with a small fraction of young stars.

\begin{figure*}
\centering
\includegraphics[width=500px]{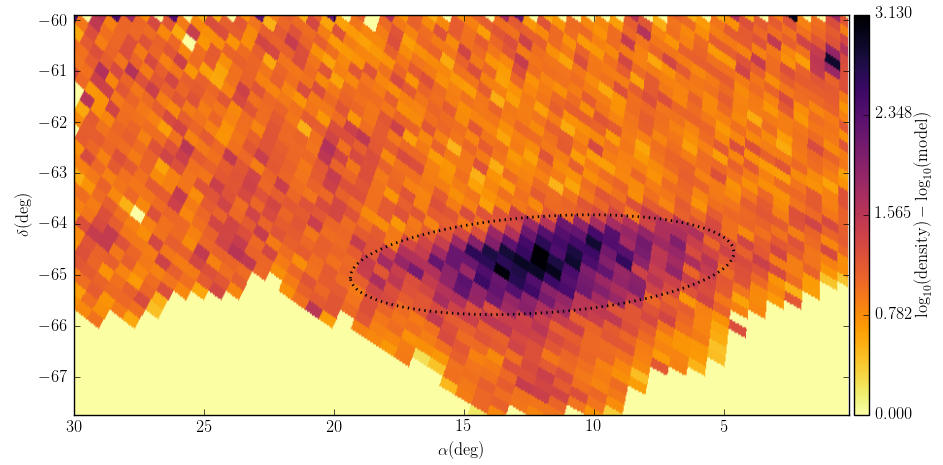}
\caption{Ratio from Fig.~\ref{DPC} (bottom panel) projected onto the sky. The truncation radius from Tab.~\ref{features} is shown as a dotted black line.}
\label{DPC2}
\end{figure*}

\begin{table}
\centering
\begin{tabular}{ l | r | c }
\hline
Property & Value & Unit \\
\hline
$\alpha$ & 12.00$^{+0.08}_{-0.06}$ & deg \\
$\delta$ & $-64.80^{+0.05}_{-0.08}$ & deg \\
$l$ & 303.53 & deg\\
$b$ & -52.32 & deg \\
$\epsilon$ & 0.60$^{+0.19}_{-0.20}$ & - \\
%$r_e$ & 100 & arcmin \\
$r_h$ & $120.4^{+19.2}_{-3.12}$ & arcmin \\
$r_{tr}$ & $192 \pm 20.0$ & arcmin \\
$M_V$ & $-7.7 \pm 0.3$ & mag \\
$\mu_V(r<r_h)$ & $31.23 \pm 0.21$ & mag arcsec$^{-2}$ \\
%$TS$ & \\
\hline
\end{tabular}
\caption{SMCNOD properties: equatorial ($\alpha$ and $\delta$) and Galactic ($l$ and $b$) coordinates of the center, ellipticity, half-light radius, truncation radius, absolute magnitude and surface brightness. The last two properties are in the $V$ band.}
\label{features}
\end{table}

The structural parameters listed in Table~\ref{features} were fit using a marginalized likelihood approach and~\textit{emcee}, an affine-invariant ensemble sampler for Markov Chain Monte Carlo models~\citep{2013PASP..125..306F}\footnote{http://dan.iel.fm/emcee/current/}. We applied the marginalized likelihood fit to the~\textsc{Healpix} pixels from Figure~\ref{DPC2}, modelling the stellar density with a Plummer profile~\citep{1911MNRAS..71..460P}.

The absolute magnitude $M_V$ was derived by adding the $V$ flux within the ellipse bounded by the SMCNOD truncation radius converted from the $g$ and $r$ bands, also using equation from~\citealt{2005AJ....130..873J}:
\begin{equation}
V = g_{SDSS} - 0.590(g_{SDSS}-r_{SDSS}) - 0.01
\end{equation}
\noindent
To evaluate the $V$ band flux of the background, we added the flux within an ellipse with the same area shifted to 3$^{\circ}$ north of the SMCNOD, then  subtracting from the SMCNOD flux in the $V$ band. The $M_V$ uncertainty incorporates the spatial fluctuations in the background flux and is in fact dominated by them.

The stellar mass of the SMCNOD was estimated by comparing a LF from a simulated simple stellar population to the SMCNOD LF. The SMCNOD LF is subtracted from a field LF immediately above the over-density, with equal area and located 3$^\circ$ north of the SMCNOD center. The subtracted LF comprises 6068 stars within the range $21.0 \leq g \leq 23.0$, corresponding to the mass range of {0.90-0.99} $M_{\odot}$. A simulated simple stellar population ($\tau$=6 Gyr and Z=0.001) with an evolved Kroupa mass function~\citep{2001MNRAS.322..231K} was generated using \textsc{gencmd}\footnote{\textsc{gencmd} yields position, magnitude and errors with a simple stellar population. See details in https://github.com/balbinot/gencmd.}, populating the {0.90-0.99} $M_{\odot}$ mass range with the same star counts as the SMCNOD subtracted LF. The mass range between {0.1-1.02} $M_{\odot}$ amounts to a stellar mass for the SMCNOD $\simeq 1.1 \times 10^5\ M_{\odot}$, and its resulting $M/L$ is very close to unity (1.07 $M_{\odot}/L_{\odot}$). The young population density is about one tenth of the intermediate-age population density and thus the computed young population mass is included in the mass error range for SMCNOD. As a comparison, the SMCNOD has a stellar mass comparable to Galactic Globular cluster NGC 6287~\citep{1997ApJ...474..223G}, but it is brighter (for NGC 6287, $M_V=-7.36$, following~\citealt[][updated 2010]{1996AJ....112.1487H}). Another estimate, using an evolved mass function similar to that found for Palomar 5  by~\citet{2004AJ....128.2274K}, where the fainter stars were removed from the main body, yields a SMCNOD stellar mass $\simeq 8.0 \times 10^4\ M_{\odot}$. These values show how the choice of a MF changes the total estimated stellar mass. Therefore, we interpret the first estimate as an upper limit for the SMCNOD stellar mass. %In a similar manner, we determine the absolute magnitude $M_V$, comparing the SMCNOD brightness to a comparison field and assuming the brightness excess in SMCNOD field is due to SMCNOD stars.

The dynamical mass of the SMCNOD ($m$) was estimated (in the case SMCNOD is bounded to the SMC) using eq. 7-84 from \cite{2008gady.book.....B}:
\begin{equation}
\label{dynmass}
\frac{m}{M} = 3 \left(\frac{r_J}{D}\right)^3
\end{equation}
\noindent
where $M$ is the SMC dynamical mass, $r_J$ is the SMCNOD tidal radius and $D$ is the distance between the SMC and SMCNOD centers. Assuming the SMCNOD tidal radius as 1.5$^{\circ}$ (from the bottom panel in Figure~\ref{DPC}) and $D$ = 8$^{\circ}$, we determined that $m/M\ \cong \ 2.0 \times 10^{-2}$. \citet{2009MNRAS.395..342B} estimate that the SMC dark halo has a mass of $3 \times 10^9$ $M_{\odot}$ in the inner 3 kpc for a $V$ band mass to light ratio $\simeq\ 2$. This mass agrees reasonably well with SMC rotation curves. Using this conservative estimate for the SMC mass, the SMCNOD dynamical mass is 6 $\times\ 10^{7}$ $M_{\odot}$, a six hundred times greater than calculated for the stellar mass. The large disagreement between the stellar and the dynamical mass calculated using Equation~\ref{dynmass} is an argument favoring the SMCNOD is a structure detached from the SMC. The uncertainties in the SMCNOD dynamical mass are dominated by the errors in the SMC dynamical mass, which is estimated as about 13\%~\citep{2009MNRAS.395..342B}. 

\begin{figure}
\centering
\includegraphics[width=280px]{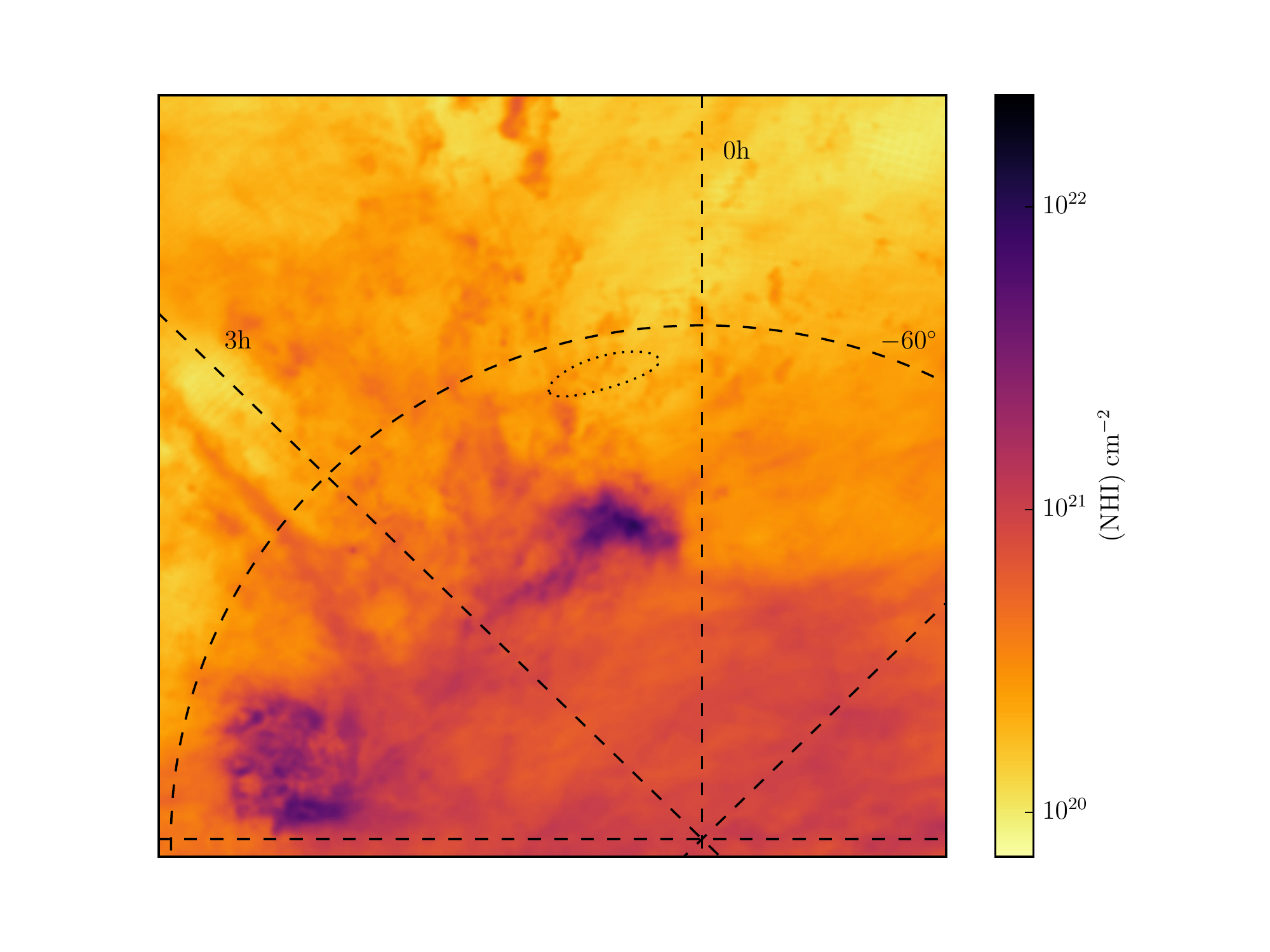}
\caption{HI gas column density map from~\citet{2015A&A...578A..78K}, in zenith equal area (ZEA) projection, close to the South Celestial Pole. A grid of equatorial coordinates ($\alpha$ and $\delta$) is indicated. The SMC is close to the center of the figure and the LMC is located near the bright spot in the lower left corner. No significant excess of HI gas is observed at the position of the SMCNOD (dotted ellipse).}
\label{HI}
\end{figure}

The HI gas map could provide more insight into the nature of the SMCNOD, as well as a possible connection to the SMC. The HI gas column density map from the GASS Third Data Release\footnote{https://www.astro.uni-bonn.de/hisurvey/gass/}~\citep{2015A&A...578A..78K} is shown in Figure~\ref{HI}. While the LMC and the SMC HI gas contents are obvious, there is no apparent excess of gas associated with the position of the SMCNOD. We also use the GASS data to look for peaks in the velocity distribution of the gas within the velocity range from -495 km/s to 495 km/s (1 km/s steps). The emission for one square degree centered on the SMCNOD exhibits two main peaks: 94 km/s and 186 km/s (which are shown in Figure~\ref{V94} and~\ref{V186}, respectively). These velocities agree with the velocity field related to the Magellanic Stream at the SMCNOD position, at a MS longitude L$_{MS} \simeq$ -25$^{\circ}$. See for example the Figure 8 in ~\citet{2010ApJ...723.1618N}. We discuss details about HI gas distribution in Sec.~\ref{discussion}.

\section{Discussion}
\label{discussion}

In this Section we discuss the characteristics of the SMCNOD: we compare its stellar populations to those of the SMC (Section~\ref{pop}), its gas content (Section~\ref{gas}), possible scenarios for its origin (Section~\ref{origin}) and in the last subsection we provide a brief summary of the discovery, discussing the SMCNOD fate and some prospects for future analyses (Section~\ref{conc}).
\subsection{The SMCNOD and SMC stellar populations}
\label{pop}

The $(g-r)$ color distributions of SMCNOD and SMC stars for three magnitude ranges in the Hess diagram from Figure~\ref{Hessdiagram} are shown in the left panel of Figure~\ref{Histo}, whereas their number counts in bins of $g$ magnitude filtered by the CMD mask and normalized (to the areas) are presented in the right panel of the same figure. The color distributions look very similar. For RC stars (solid lines), there may be a slight preference for the SMCNOD being a little bluer than the SMC. However, a Kolmogorov-Smirnov (KS) test indicates that the two RC populations come from the same parent distribution ($p = 0.42$). As for the luminosity function comparison, the two distributions look similar as well.

\subsection{The SMCNOD and HI gas}
\label{gas}
\begin{figure}
\centering
\includegraphics[width=280px]{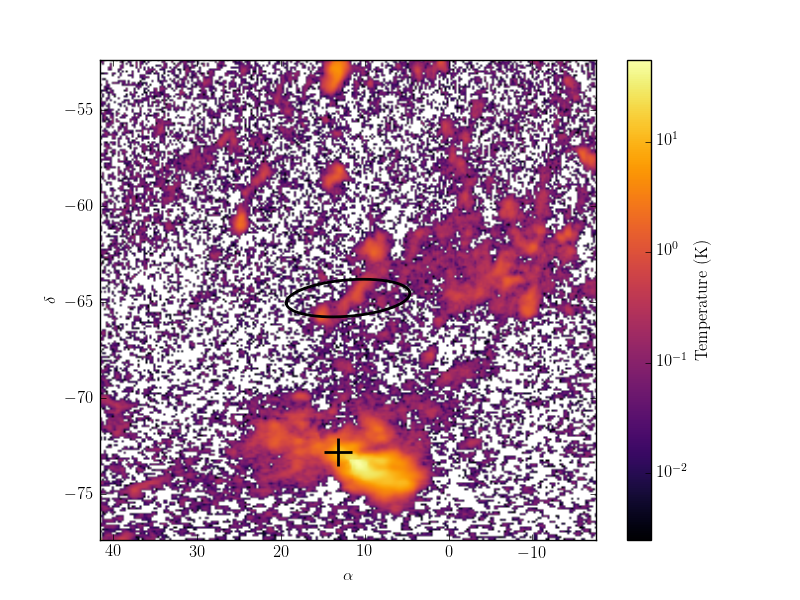}
\caption{HI gas emission map from~\citet{2015A&A...578A..78K}, in equatorial coordinates for the velocity channel 94 km s$^{-1}$ $<$ v$_{LSR}$ $<$ 95 km s$^{-1}$. The SMCNOD position is highlighted by an empty ellipse and the SMC center by a plus symbol.}
\label{V94}
\end{figure}

\begin{figure}
\centering
\includegraphics[width=280px]{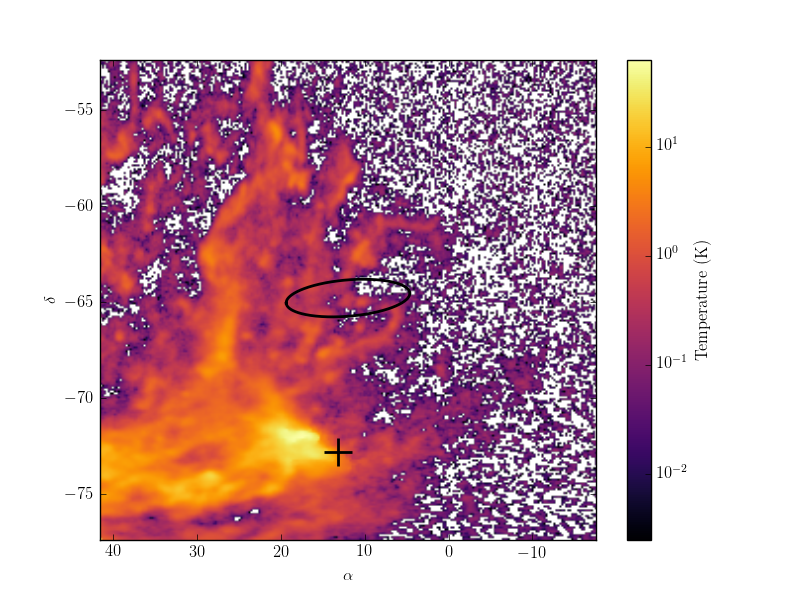}
\caption{Same as Fig.~\ref{V94} but for velocity channel equal to 186 km s$^{-1}$ (186 km s$^{-1}$ $<$ v$_{LSR}$ $<$ 187 km s$^{-1}$).}
\label{V186}
\end{figure}

\begin{figure*}
\centering
\includegraphics[width=520px]{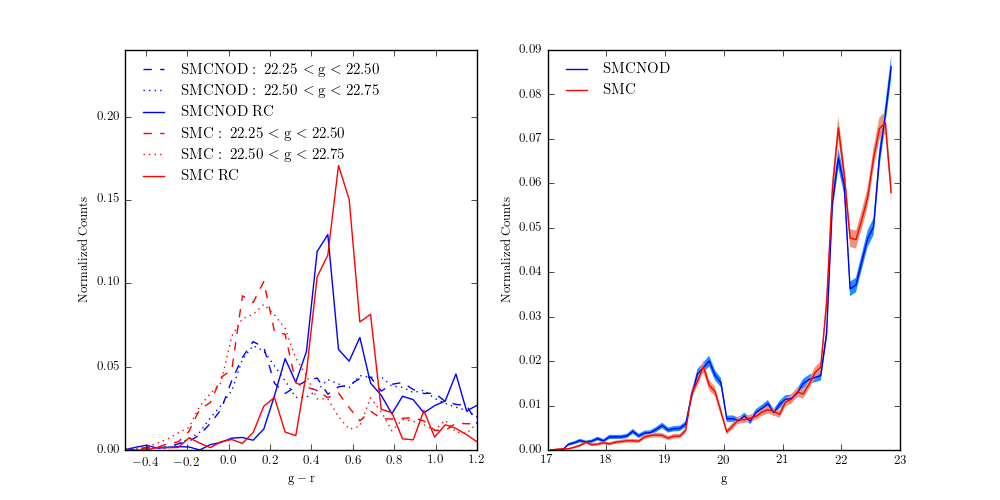}
\caption{Histograms in bins of color (left) and in bins of $g$ magnitude (right) filtered by the CMD mask for the SMCNOD (third panel in Fig.~\ref{Hessdiagram}) and the SMC (fifth panel in Fig.~\ref{Hessdiagram}). In the right panel, the Poissonian uncertainties are shown as shaded areas. The stellar color distribution is very similar between the SMC and SMCNOD in the left panel (comparing dotted red and blue lines and dashed red and blue lines), as well as the histogram in bins of magnitude for the stars filtered with the CMD mask in the last two panels in Fig.\ref{Hessdiagram} (right panel).}
\label{Histo}
\end{figure*}

The results of considering the HI gas velocity channels are inconclusive. The v$_{LSR}$=186 km/s channel map (Figure~\ref{V186}) shows a few links between the SMC and the SMCNOD, while the v$_{LSR}$=94 km/s channel map (Figure~\ref{V94}) shows a bar-shaped gas cloud detached from the SMC main body. A looping feature is visible in the v$_{LSR}$=186 km/s channel map that could be the result of a weak gas inflow (from the SMC, counterclockwise). But in summary, the SMCNOD does not seem to contain a large amount of gas and it is currently unclear whether the HI gas features present in either velocity channel map are connected to it. It has been suggested that the drift rate of the MS gas away from the LMC is $\sim$ 49 km/s, as indicated by~\citet{2008ApJ...679..432N}. Using these results, an age of 1.74 Gyr is expected for the MS. In this sense, the gas features surrounding the SMC should be very recent (a few hundred million years or even younger), showing a complex dynamics. Taken at face value, the gas properties around the SMCNOD are more consistent with gas-poor dwarf spheroidals~\citep{2003AJ....125.1926G} and ultra-faint dwarf galaxies in the Local Group~\citep{2013PhDT........14G}.

\subsection{The SMCNOD origin}
\label{origin}
Regarding the formation of the SMCNOD, the most likely scenario is that this structure was formed by material pulled from the SMC disk through tidal stripping, given the recent N-body simulations as cited in Sec.~\ref{intro}. Following the classical galactic interaction theory of \citet{1972ApJ...178..623T}, many other works predict the existence of a Magellanic Counter-Bridge as a counterpart to the Magellanic Bridge \citep{2014MNRAS.442.1663D,2012ApJ...750...36D}. In the north-west part of the SMC a kinematical substructure discovered by \citet{2014MNRAS.442.1663D} is associated to the Magellanic Counter-Bridge, as an observational counterpart. 
The simulations also predict a spread of stars as a result of an LMC-SMC close encounters. See for example the tidal tail in Figure 5 of~\citet{1996MNRAS.278..191G}, the set of particles located southwest of the SMC in Figure 12 from \citet{2003MNRAS.339.1135Y}, and the SMC stellar distribution in Figure 4 and 5 of \citet{2006MNRAS.371..108C}. Also, a conspicuous clump of young stars can be seen for the fiducial model simulated by \citet{2009PASA...26...48B} in their Figure 5, 6 and 9, located 4 degrees from the center of the SMC, along with a stream-like feature on the opposite side. Earlier simulations present a spread of particles around the SMC, for example Figure 10b of \citet{1976A&A....47..263F} and Figure 6a of \citet{1980PASJ...32..581M}. The substantial stream-like stellar overdensity in the northern periphery of the LMC centre recently discovered by~\citet{2016MNRAS.459..239M}, with characteristics similar to the SMCNOD, also corroborates this scenario {for the SMCNOD formation} based on close LMC-SMC interactions. Finally, a close encounter occurred $\approx$ 200 Myr ago is also claimed as an explanation for a 55 kpc stellar structure in the eastern SMC \citep{2013ApJ...779..145N}, where likely the stars where tidally stripped from the SMC.

If the SMCNOD is the result of an LMC-SMC collision, a contemporaneous peak in star formation is expected in both galaxies. Unfortunately, the results of star formation history (SFH) analyses have large uncertainties, and there is a significant spatial variation for the SFH in the LMC~\citep{2012A&A...537A.106R,2009AJ....138.1243H,2002ApJ...566..239S,1999AJ....117.2244O,1999AJ....118.2262H} and SMC~\citep{2015MNRAS.449..639R,2013ApJ...775...83C,2009ApJ...705.1260N}. SFH variations are larger when based on different models (isochrones) and/or stellar tracers (RGB, carbon or variable stars) and the MCs SFH is still far from being fully characterized in spite of much work. Even so, it is interesting to note that most of the references listed above agree with a simultaneous peak in SFR between 4-6 Gyr (also discussed in \citealt{2014MNRAS.442.1680D}), the age of the main stellar population of the SMCNOD. A complete reconstruction of the SFH of the SMCNOD (and also a comparison to the SFH of various SMC regions) is beyond the scope of this work.

Another possible scenario for the SMCNOD origin is a tidal dwarf galaxy (TDG), an object formed as described by~\citet{1993ApJ...412...90E}, where mainly gas is stripped from past mergers and resemble as dwarf galaxies, where the stars are forming during/after the main encounter. The lack of dark matter in TDGs makes them fragile, leading to short lifetimes (a few Gyr) as cohesive systems. This scenario for the SMCNOD formation is disfavored due to its poor gas content and predominantly intermediate-age stars (at least 6 Gyr), compared to an expected young TDG stellar population~\citep{2012ASSP...28..305D}. On the other hand, numerical simulations by~\citet{2014MNRAS.437.3980P} show TDGs could survive at least 3 Gyr, despite the lack of dark matter content. This is corroborated by the existence of the relatively old TDG VCC2062 observed by~\citet{2007A&A...475..187D}, where its parent galaxies have likely merged.

The scenario where the SMCNOD is a primordial galaxy orbiting and/or merging with the SMC could be favored if the stellar populations of the SMCNOD have narrower age and metallicity ranges than those of the SMC. This would make the SMCNOD more consistent with typical dwarf spheroidal galaxies. 

An intriguing possibility for the origin of the SMCNOD is the ``resonant stripping'' predicted by \cite{2009Natur.460..605D}. This process allows for an efficient removal of stars in a dwarf-dwarf encounter, where the smaller dwarf loses stars by a resonance between the angular frequency of its orbit and spin, changing the ratio of the stellar to dark matter mass. Simulations also predict that dwarf-disk galaxies will evolve into compact spheroidal systems with stream-like and shell-like structures, resembling the SMCNOD shape around the SMC. Future deep photometric surveys of the SMC/LMC outskirts could reveal similar structures, testing the significance of this ``resonant stripping'' in the model.

The relatively old age for most of the stars in the SMC-NOD rules out its origin as being formed by HI gas from the Magellanic Stream. As~\citet{2008ApJ...679..432N} point out, the expected age of putative MS stars should be $\simeq$ 2.0 Gyr, less than half of the characteristic age found for the over-density stars. The origin of the SMCNOD younger population may or may not be attributed to the same physical mechanism as the intermediate-age population. Since the SMC-LMC interaction is known to have triggered star formation at recent times (as in the case of the Magellanic Bridge, see for example~\citealt{2015MNRAS.453.3190B} and~\citealt{2015MNRAS.452.4222N}), we cannot rule out that this interaction may be responsible for the younger SMCNOD population.

\subsection{Summary, prospects and fate of the SMCNOD}
\label{conc}
Using DES and MagLiteS data, we have found and analyzed a stellar over-density located about 8$^{\circ}$ north of the SMC center. The stellar density and surface brightness associated with this feature lie significantly above the values expected from the extrapolated stellar profile of the SMC itself. This is true even when we consider only the contribution from the RGB stars. Previous surveys around the SMCNOD , such as the Magellanic Periphery Survey \citep{2011ApJ...733L..10N}, Two Micron All Sky Survey \citep{1997ASSL..210...25S}, Digitized Sky Survey\footnote{http://archive.stsci.edu/cgi-bin/dss\_form} and Infrared Astronomical Satellite survey\footnote{http://irsa.ipac.caltech.edu/Missions/iras.html}, among others, did not reveal any stellar over-density similar to the one measured here. This may be due to either their non-contiguous area or their lower photometric depth.

The fact that the structure~discussed here has a density peak lying significantly in excess of the expected SMC exponential density profile (or above the combined SMC and Galactic background) and follows a roughly elliptical profile encourage us to argue that it is a distinct SMC substructure. On the other hand, the CMD analysis indicates that the stellar populations are very similar to those found in the main SMC body, and the SMCNOD lies at a similar heliocentric distance as the SMC.

The fate of the SMCNOD has interesting implications for the Magellanic System generally. If the SMCNOD is an unstable object, such as a stellar cloud, it should dissipate into the SMC main body or, if unbound to the SMC, be ejected and dissipate eventually into the Galactic field. However, for the first hypothesis to hold, the SMC truncation radius must be$~\simeq 10$ kpc (see the SMCNOD limits in Figure~\ref{DPC}) to encompass the entire stellar cloud presented here. {As a reference,} the truncation radius derived from chemo-dynamical simulations involving SMC-like objects is in the range between 5-7.5 kpc~\citep{2009PASA...26...48B,2007PASA...24...21B,2012ApJ...750...36D}, not enough to include the entire SMCNOD.

Radial velocities and proper motions of likely stellar members will constrain systemic and internal kinematics of the SMCNOD, as well as its internal motions. Metallicities and other abundance estimates may indicate similarities and differences between the SMCNOD stars and those belonging to the main SMC body. An internal age and/or metallicity gradient (or its absence) may also constrain its nature as either a primordial or tidal object. Finally, the SMCNOD discovery shows that the Magellanic System, despite being relatively well-studied, still hides surprising substructures that may be revealed with deep photometric surveys. The discovery of the SMCNOD at such a large distance from the SMC should provide an additional constraint for simulations of the Magellanic System.

\section*{Acknowledgments}

We are grateful to the anonymous referee for valuable comments, which contributed to improving this paper.
This paper has gone through internal review by the DES and MagLiteS Collaboration. EB acknowledges financial support from the European Research Council (ERC-StG-335936). DMD acknowledges support by Sonderforschungsbereich (SFB) 881 ``The MilkyWay System'' of the German Research Foundation (DFB), subproject A2.

Funding for the DES Projects has been provided by the U.S. Department of Energy, the U.S. National Science Foundation, the Ministry of Science and Education of Spain, the Science and Technology Facilities Council of the United Kingdom, the Higher Education Funding Council for England, the National Center for Supercomputing Applications at the University of Illinois at Urbana-Champaign, the Kavli Institute of Cosmological Physics at the University of Chicago, the Center for Cosmology and Astro-Particle Physics at the Ohio State University, the Mitchell Institute for Fundamental Physics and Astronomy at Texas A\&M University, Financiadora de Estudos e Projetos, Funda{\c c}{\~a}o Carlos Chagas Filho de Amparo {\`a} Pesquisa do Estado do Rio de Janeiro, Conselho Nacional de Desenvolvimento Cient{\'i}fico e Tecnol{\'o}gico and the Minist{\'e}rio da Ci{\^e}ncia, Tecnologia e Inova{\c c}{\~a}o, the Deutsche Forschungsgemeinschaft and the Collaborating Institutions in the Dark Energy Survey. 

The Collaborating Institutions are Argonne National Laboratory, the University of California at Santa Cruz, the University of Cambridge, Centro de Investigaciones Energ{\'e}ticas, Medioambientales y Tecnol{\'o}gicas-Madrid, the University of Chicago, University College London, the DES-Brazil Consortium, the University of Edinburgh, the Eidgen{\"o}ssische Technische Hochschule (ETH) Z{\"u}rich, Fermi National Accelerator Laboratory, the University of Illinois at Urbana-Champaign, the Institut de Ci{\`e}ncies de l'Espai (IEEC/CSIC), the Institut de F{\'i}sica d'Altes Energies, Lawrence Berkeley National Laboratory, the Ludwig-Maximilians Universit{\"a}t M{\"u}nchen and the associated Excellence Cluster Universe, the University of Michigan, the National Optical Astronomy Observatory, the University of Nottingham, The Ohio State University, the University of Pennsylvania, the University of Portsmouth, SLAC National Accelerator Laboratory, Stanford University, the University of Sussex, Texas A\&M University, and the OzDES Membership Consortium.

The DES data management system is supported by the National Science Foundation under Grant Number AST-1138766.
The DES participants from Spanish institutions are partially supported by MINECO under grants AYA2012-39559, ESP2013-48274, FPA2013-47986, and Centro de Excelencia Severo Ochoa SEV-2012-0234.
Research leading to these results has received funding from the European Research Council under the European Union’s Seventh Framework Programme (FP7/2007-2013) including ERC grant agreements 240672, 291329, and 306478.
%\nocite{*}
\bibliographystyle{mn2e}

%\bibliography{}
{}

\section*{Affiliations}
{\small\it
\noindent
$^1$ Instituto de F\'{\i}sica, Universidade Federal do Rio Grande do Sul, 91501-900 Porto Alegre, RS, Brazil \\
$^2$ Laborat{\' o}rio Interinstitucional de e-Astronomia - LIneA, Rua Gal. Jos{\' e} Cristino 77, 20921-400, \\ Rio de Janeiro, RJ, Brazil \\
$^3$ Fermi National Accelerator Laboratory, P. O. Box 500, Batavia, IL 60510, USA \\
$^{4}$ Large Synoptic Survey Telescope, 933 North Cherry Avenue, Tucson, AZ 85721, USA \\
$^5$ Space Telescope Science Institute, 3700 San Martin Drive, Baltimore, MD 21218, USA \\
$^6$ Steward Observatory, University of Arizona, 933 North Cherry Avenue, Tucson, AZ, 85721, USA \\
$^{7}$ Observatoire astronomique de Strasbourg, Universit{\'e} de Strasbourg, CNRS, UMR 7550, 11 rue de l'Universit{\'e}, F-67000 Strasbourg, France \\
$^{8}$ Max-Planck-Institut f{\"u}r Astronomie, K{\"o}nigstuhl 17, D-69117 Heidelberg, Germany \\
$^{9}$ Institute of Astronomy, University of Cambridge, Madingley Road, Cambridge CB3 0HA, UK \\
$^{10}$ Instituto de Astrof{\'i}sica de Canarias. V{\'i}a L{\'a}ctea s/n. E38200 - La Laguna, Tenerife, Canary Islands, Spain \\
$^{11}$ Department of Astrophysics, University of La Laguna. V{\'i}a L{\'a}ctea s/n. E38200 - La Laguna, Tenerife, Canary Islands, Spain \\
$^{12}$ Astronomisches Rechen-Institut, Zentrum f{\"u}r Astronomie der Universit{\"a}t Heidelberg, M{\"o}nchhofstr. 12-14, D-69120 Heidelberg, Germany \\
$^{13}$ George P. and Cynthia Woods Mitchell Institute for Fundamental Physics and Astronomy, and Department of Physics and Astronomy, Texas A\&M University, College Station, TX 77843, USA \\
$^{14}$ Department of Physics, University of Surrey, Guildford GU2 7XH, UK \\
$^{15}$ Department of Astronomy, University of Virginia, Charlottesville, VA 22904-4325, USA \\
$^{16}$ University of Hertfordshire, Physics Astronomy and Mathematics, College Lane, Hatfield AL10 9AB, United Kingdom \\
$^{17}$ Leibnitz-Institut f{\"u}r Astrophysik Potsdam, An der Sternwarte 16, D-14482 Potsdam, Germany \\
$^{18}$ Department of Physics, ETH Zurich, Wolfgang-PauliStrasse 16, CH-8093 Zurich, Switzerland \\
$^{19}$ Research School of Astronomy \& Astrophysics, Mount Stromlo Observatory, Cotter Road, Weston Creek, ACT 2611, Australia \\
$^{20}$ Cerro Tololo Inter-American Observatory, National Optical Astronomy Observatory, \\ Casilla 603, La Serena, Chile \\
$^{21}$ Center for Astrophysics and Space Astronomy, Department of Astrophysical and Planetary Sciences, University of Colorado, 389 UCB, Boulder, CO 80309, USA \\
$^{22}$ National Optical Astronomy Observatory, 950 N. Cherry Ave., Tucson, AZ 85719, USA \\
$^{23}$ Observat{\' o}rio Nacional, Rua Gal. Jos{\' e} Cristino 77, Rio de Janeiro, RJ - 20921-400, Brazil  \\
$^{24}$ Department of Physics and Electronics, Rhodes University, PO Box 94, Grahamstown, 6140, South Africa \\
$^{25}$ Department of Physics \& Astronomy, University College London, Gower Street, London, WC1E 6BT, UK \\
$^{26}$ CNRS, UMR 7095, Institut d'Astrophysique de Paris, F-75014, Paris, France \\
$^{27}$ Sorbonne Universit\'es, UPMC Univ Paris 06, UMR 7095, Institut d'Astrophysique de Paris, \\ F-75014, Paris, France \\
$^{28}$ Department of Astronomy, University of Illinois, 1002 W. Green Street, Urbana, IL 61801, USA \\
$^{29}$ National Center for Supercomputing Applications, 1205 West Clark St., Urbana, IL 61801, USA \\
$^{30}$ Institut de Ci\`encies de l'Espai, IEEC-CSIC, Campus UAB, Carrer de Can Magrans, s/n, 08193 \\ Bellaterra, Barcelona, Spain \\
$^{31}$ Instituto de Fisica Teorica UAM/CSIC, Universidad Autonoma de Madrid, 28049 Madrid, Spain \\
$^{32}$ Kavli Institute for Particle Astrophysics \& Cosmology, P. O. Box 2450, Stanford University, Stanford, CA 94305, USA \\
$^{33}$ Institute of Cosmology \& Gravitation, University of Portsmouth, Portsmouth, PO1 3FX, UK \\
$^{34}$ School of Physics and Astronomy, University of Southampton,  Southampton, SO17 1BJ, UK \\
$^{35}$ Department of Physics, IIT Hyderabad, Kandi, Telangana 502285, India \\ 
$^{36}$ SLAC National Accelerator Laboratory, Menlo Park, CA 94025, USA \\
$^{37}$ Center for Cosmology and Astro-Particle Physics, The Ohio State University, Columbus, OH 43210, USA \\
$^{38}$ Department of Physics, The Ohio State University, Columbus, OH 43210, USA \\
$^{39}$ Astronomy Department, University of Washington, Box 351580, Seattle, WA 98195, USA \\
$^{40}$ Australian Astronomical Observatory, North Ryde, NSW 2113, Australia \\ 
$^{41}$ Instituci\'o Catalana de Recerca i Estudis Avan\c{c}ats, E-08010 Barcelona, Spain \\
$^{42}$ Institut de F\'{\i}sica d'Altes Energies (IFAE), The Barcelona Institute of Science and Technology, Campus UAB, 08193 Bellaterra (Barcelona) Spain \\
$^{43}$ Jet Propulsion Laboratory, California Institute of Technology, 4800 Oak Grove Dr., Pasadena, CA 91109, USA
$^{44}$ Department of Physics and Astronomy, Pevensey Building, University of Sussex, Brighton, BN1 9QH, UK \\
$^{45}$ Department of Physics and Astronomy, University of Pennsylvania, Philadelphia, PA 19104, USA
$^{46}$ Centro de Investigaciones Energ\'eticas, Medioambientales y Tecnol\'ogicas (CIEMAT), Madrid, Spain \\
$^{47}$ Department of Physics, University of Michigan, Ann Arbor, MI 48109, USA \\
$^{48}$ Universidade Federal do ABC, Centro de Ci\^encias Naturais e Humanas, Av. dos Estados, 5001, Santo Andr\'e, SP, Brazil, 09210-580 \\
$^{49}$ Computer Science and Mathematics Division, Oak Ridge National Laboratory, Oak Ridge, TN 37831}

\label{lastpage}

\end{document}